\documentclass[twocolumn,showpacs,preprintnumbers,amsmath,amssymb]{revtex4}

\usepackage{graphicx}
\usepackage{dcolumn}
\usepackage{bm}
\usepackage{amssymb}
\usepackage{enumerate}

\newcommand\ee{e^+e^-}

\newcommand\zm{Z_\mu}
\newcommand\zmp{Z'}
\newcommand\react{\mu + Z \to \mu + Z + \zm}
\newcommand\mm{\mu^+ \mu^- }

\newcommand\zmvis{Z_\mu \to \mu^+ \mu^-}
\newcommand\zmnn{Z_\mu \to \nu \nu}

\def\address{\@ifstar{\address@star}%
  {\@ifnextchar[{\address@optarg}{\address@noptarg}}}

\begin{document}
\author{S.N.~Gninenko$^{1}$}
\author{N.V.~Krasnikov$^{1,2}$}
\author{V.A.~Matveev$^{1,2}$}

\affiliation{$^{1}$ Institute for Nuclear Research of the Russian Academy of Sciences, 117312 Moscow, Russia \\
$^{2}$ Joint Institute for Nuclear Research, 141980 Dubna, Russia}


\title{Muon g-2 and searches for  a new  leptophobic sub-GeV dark  boson in a missing-energy experiment at CERN  } 

\date{\today}

\begin{abstract}

The 3.6 $\sigma$ discrepancy between the predicted and  measured values of the 
anomalous  magnetic moment of positive muons can be explained by  
the existence of a new dark boson $\zm$ with a mass in the sub-GeV range, which is coupled predominantly to the second and third lepton generations through the $L_\mu - L_\tau$ current. After  a discussion of the present phenomenological bounds on the $\zm$ coupling, we show that if the $\zm$  exists, it could be observed in the reaction $\react$ of a muon scattering off nuclei  by looking for an excess of events with large missing muon beam energy in a detector due to the prompt bremsstrahlung $\zm$ decay $\zmnn$ 
into a couple of neutrinos. We describe the experimental technique and
the preliminary  study of the feasibility for the proposed search. We show  that this specific signal allows for a search for the $\zm$ with a sensitivity in the coupling constant $\alpha_\mu \gtrsim 10^{-11}$, which is 3 orders of magnitude higher than the value required to explain the discrepancy. We point out  
that the availability of high-energy and -intensity muon beams at CERN SPS provides 
unique opportunity to either discover or rule out the $\zm$ in the proposed search in the near future.  
 The experiment is based on the missing-energy approach developed for the searches for
invisible decays of dark photons and (pseudo)scalar mesons at CERN  and is complementary to these experiments.
\end{abstract}
\maketitle

\section{Introduction}
The precise measurement of the  anomalous magnetic 
moment of the positive muon $a _{\mu}= (g-2)/2$ from the Brookhaven AGS experiment 821 \cite{bnl} gives a result which is about
$3.6 \sigma$ higher than the 
Standard Model (SM) prediction 
\begin{equation}
a^{exp}_{\mu} -a^{SM}_{\mu} = 288(80)\times 10^{-11}
\label{bnl}
\end{equation}
This result may signal the existence of new physics beyond the 
Standard Model. 
At present the most popular   explanation of this discrepancy   is supersymmetry with 
a chargino and sneutrino lighter than 800 GeV \cite{theor2}. 
Other  possible explanations include  leptoquarks 
\cite{theor3} 
or  some exotic flavor-changing interactions \cite{theor4}. 
All of these explanations assume the existence of new heavy particles with masses 
$\geq O(100)~GeV$. Another  explanation of the $g-2$ anomaly is related to the existence of a new light  ( with a mass $m_{Z'} \leq O(1)$ GeV)
vector boson (dark photon) which couples very weakly with the muon with 
$\alpha_{Z'} \sim O(10^{-8})$  \cite{theor5} -\cite{theor10}, see also Ref. \cite{theor11}. 

In this paper we consider  the muon $g-2$ anomaly as an indication for the 
existence of the  new light vector boson $\zm$, which is coupled predominantly to the second and third lepton generations. We propose an experiment  to search for the $\zm$ 
in the high-energy muon  beam at the CERN SPS.  
If the $\zm$  exists, it could be observed in the reaction $\react$
of a high-energy muon scattering off 
nuclei  by looking for an excess of events with 
a specific  signature, namely large missing  muon beam energy in the detector.  The experiment uses the missing-energy approach developed for the search for invisible decays of dark photons and (pseudo)scalar mesons at CERN \cite{sngldms,ldms,snginv} and is complementary to these  proposals.     

 The rest of the paper is organized in the following way. 
The existing bounds are discussed in Sec.II. In Sec. III the $Z_{\mu}$ production and decay modes are described. 
The method of the search and the experimental setup are presented in Sec. IV, background sources are discussed in Sec. V, and   the expected sensitivity   is shown in Sec. VI. Section VII contains concluding remarks. 
  
\section{Phenomenology and existing experimental bounds }
As discussed in the Introduction one of the possible explanations of the $g_{\mu} - 2$ anomaly assumes the existence 
of a new light vector boson $\zmp$ which interacts with muons like a photon, namely 
\begin{equation}
L_{\zmp} = e'\bar{\mu}\gamma_{\nu}\mu \zmp^\nu \,.
\label{int}
\end{equation}
The interaction (2) gives additional contribution to the muon anomalous 
magnetic moment $a_{\mu} \equiv \frac{g_{\mu} - 2}{2}$
\begin{equation}
a'_l = \frac{\alpha'}{\pi}\int^1_0 \frac{x^2(1-x)}
{x^2 + (1-x)M^2_{\zmp}/m^2_l} \,,
\end{equation}       
where $\alpha' = (e')^2 /4\pi $ and $M_{\zmp}$ is the mass of the $\zmp$ boson. 
Equation (3) allows one to determine the $\alpha_\mu$ which explains  the $g_{\mu} - 2$ anomaly.
For 
$M_{\zmp}\ll m_{\mu}$ we find from Eq.(1) that
\begin{equation} 
\alpha' =  (1.8 \pm 0.5) \times 10^{-8} 
\end{equation}
For another limiting case $M_{\zmp} \gg m_{\mu}$  Eq.(1) leads to  
\begin{equation}
\alpha' \frac {m^2_{\mu}}{M^2_{\zmp}} = (2.7 \pm 0.8)\times 10^{-8} 
\end{equation}

\begin{figure}[tbh!]
\begin{center}
\includegraphics[width=0.4\textwidth]{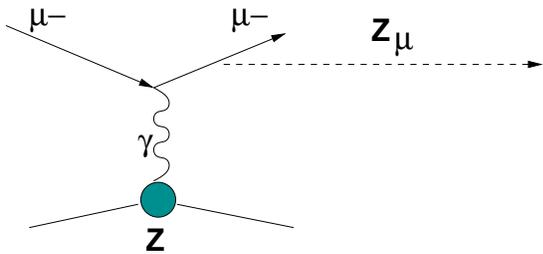}
\caption{Diagram illustrating the massive $\zm$ production in the  reaction $\react$ of muons scattering off a nuclei (A,Z). The $\zm$ is either stable or it decays invisibly if its mass $M_{\zm}\leq 2 m_\mu$, or (as shown)  it could  subsequently decay  into a $\mu^+\mu^-$ pair if $M_{\zm} > 2 m_\mu$.}
 \label{diagr}
\end{center}

\end{figure}

But the postulation of the interaction \eqref{int} of  the $\zmp$ boson with a muon  is not the end of the story. 
The main question is,  what about the interaction of the  $\zmp$  boson with other quarks and leptons?  
There are a lot of possibilities here. For instance, a very popular scenario involves an interaction of the 
 $\zmp$     boson with quarks and leptons that is proportional to the electromagnetic current $J_{\nu}$ of the SM, 
namely
\begin{equation}
L_{\zmp} = e' J_{\nu}\zmp^{\nu} \,.
\label{intmu}
\end{equation}
The natural realization of this scenario is the existence of a  new  gauge 
  boson $\zmp$ which interacts 
with the SM fields through the mixing with the SM hypercharge \cite{X1,X2}
\begin{equation}
\Delta L = \frac{\epsilon}{2}F^{Y,\alpha\beta}F^{\zmp}_{\alpha\beta} \,.
\end{equation}
 
For the scenario with the interaction \eqref{intmu} of the $\zmp$ boson with the electromagnetic current of the SM
there are several interesting constraints.
Bounds \cite{e1,e2} from the electron magnetic moment value
\begin{equation}
\Delta a_e = a^{exp}_e - a_e^{SM} = -1.06(0.82) \times 10^{-12} \,
\end{equation} 
exclude the region $M_{\zmp} < 30$ MeV.
Other experiments use dilepton resonance searches, $Z_{\mu} \rightarrow l^+l^-$. First we consider the bounds obtained under the assumption that the $\zmp$ boson decays mainly into 
charged leptons, i.e. Br$(\zmp \rightarrow l^+l^-)  = 1$, with  $l = e, \mu$. 
The Phenix Collaboration looked for the  $\zmp$ boson in $\pi^0, ~\eta \rightarrow (\zmp \rightarrow e^+e^-)\gamma$ decays 
and excluded the masses $36 < M_{\zmp} < 90$ MeV \cite{PHENIX}. The A1 Collaboration used the reaction 
$eZ \rightarrow e\zmp Z; \zmp \rightarrow e^+e^-$ to search for the $\zmp$ boson and excluded the masses 
$40 < M_{\zmp} < 300$ MeV \cite{MAMI}. The $BABAR$  Collaboration  looked for the  $Z_{\mu}$ boson in the reaction 
$e^+e^- \rightarrow \gamma\zmp, \zmp \rightarrow e^+e^-, \mu^+\mu^-$ and excluded the masses 
30 MeV $ < M_{\zmp} < 10.2$ GeV \cite{BABAR}. Finally, taking into account 
the recent results from $K-$decay experiments \cite{na48},  the possibility of the $g-2$ explanation in the model  with the interaction (6) of the $\zmp$ boson 
with the assumption that    Br$(\zmp \rightarrow l^+l^-)  = 1$ is  excluded, see, e.g. Ref.\cite{kahn} for a discussion.

For the model with the interaction \eqref{intmu} there is the possibility that the  $\zmp$ boson 
decays dominantly invisibly into new light particles $\chi$ with the branching 
Br$(\zmp \rightarrow \chi \tilde{\chi}) = 1$.   For this scenario the $K^+ \rightarrow \pi^+  ~+~ missing~energy$ 
bound \cite{BNL} and the off-resonance $BABAR$ result \cite{BABAR0} exclude a sizable parameter space, except for $30 < M_{\zmp}  < 50$ MeV and the narrow region around $M_{\zmp} = 140$ MeV \cite{BABAR1}, \cite{LEE}. 

Another interesting scenario is that from 
Ref. \cite{LEE}, where the light gauge boson 
(the dark leptonic gauge boson)  
interacts with the leptonic current, namely 
\begin{eqnarray}
L_{\zmp} = e' [\bar{e}\gamma_{\nu}e + \bar{\nu}_{eL} \gamma_{\nu}\nu_{eL} + \bar{\mu}\gamma_{\nu}\mu +  \bar{\nu}_{\mu L}\gamma_{\nu}\nu_{\mu L} \nonumber \\
+ \bar{\tau}\gamma_{\nu}\tau +  \bar{\nu}_{\tau L} \gamma_{\nu}\nu_{\tau L}]\zmp^\nu
\end{eqnarray}

This interaction does not contain quarks and as a consequence the corresponding model escapes  many quarkonium-decay constraints \cite{LEE}. 
The relevant searches for the dark leptonic gauge boson  include fixed-target \cite{MAMI} and neutrino trident experiments \cite{CHARM, CCFR}, the $BABAR$  search for $e^+e^- \rightarrow \gamma + missing~energy$ \cite{BABAR},
beam-dump experiments \cite{e137, e141, e774},  and last but not least the Borexino experiment \cite{BOREXINO}. 
It appears that for the model with the dark leptonic gauge boson the most restrictive bound comes from the  Borexino experiment \cite{BOREXINO}. 
The presence of a new vector boson would alter the charged-current interaction between solar $\nu_e$ neutrinos and target electrons in 
the detector.  The bounds from the 862 KeV $^7$Be solar neutrino flux measurement at the Borexino experiment excludes 
the possibility that the leptonic gauge boson can explain the $g_{\mu} - 2$ anomaly \cite{EXCLUSION}, 
see Fig. 3 in Ref.\cite{LEE} where constraints on the parameter space of the dark lepton gauge boson model were presented \footnote{For instance, for $m_{\zmp} = 1$ MeV the experimental bounds on $e'$ is $e' \leq 3 \times 10^{-6}$ \cite{EXCLUSION}. 
Note that in the early papers \cite{eric}-\cite{gnin} (see also Ref.\cite{okun}) similar bounds on the $\zm$-boson coupling from neutrino reactions were obtained.}.

In Refs. \cite{theor5} - \cite{theor7}, an explanation of the $g_{\mu} - 2$ anomaly was given by a model where the new light gauge boson ( hereafter denoted as  $Z_{\mu}$)  interacts with 
the $L_{\mu} - L_{\tau}$ current as   
\begin{equation}
L_{Z_\mu} = e_\mu [\bar{\mu}\gamma_{\nu}\mu +  \bar{\nu}_{\mu L} \gamma_{\nu}\nu_{\mu L}
- \bar{\tau}\gamma_{\nu}\tau  -  \bar{\nu}_{\tau L} \gamma_{\nu}\nu_{\tau L}]Z^\nu_{\mu}
\end{equation}
which  is anomaly free and  corresponds to the global flavor symmetry $U(1)_{L_\mu - L_\tau}$ which  commutes with with the SM $SU_c(3) \otimes SU(2)_L \otimes U(1)_Y$ gauge group \cite{foot}. In addition, it was recently shown that the 
$\zm$ with a mass $\simeq$ 2 MeV can explain the gap in the cosmic neutrino 
spectrum observed by the IceCube Collaboration \cite{kaneko}. 

  As the  $Z_{\mu}$ does not couple to  quarks, electrons and  $\nu_{e}$ 
neutrinos, it escapes the most current experimental constraints. The most 
restrictive bound comes from the results of experiments on neutrino trident production $\nu_{\mu}N \rightarrow \nu_{\mu}N + \mu^+ \mu^-$  \cite{CHARM, CCFR}. As was  shown in Ref. \cite{POSPELOV} 
that the CCFR data  \cite{CCFR}  on    $\nu_{\mu}N \rightarrow \nu_{\mu}N + \mu^+ \mu^-$  production exclude the $g_{\mu} -2$ explanation for a $Z_{\mu}$-boson mass $m_{Z_{\mu}} \geq 400$ MeV. 

We note that at the one-loop level the $\zm$ and the photon are kinetically 
mixed. The effective coupling of $Z_{\mu}$ to electrons (or quarks) due to the muon or $\tau$-lepton loop  is $\simeq O(\frac{\alpha}{\pi})e_{\mu}$, i.e. it is suppressed by at least a factor $ \approx  3 \cdot 10^{-3}$.  
This results in rather modest   constraints  on  invisible decays of $Z_{\mu}$ 
which one can extract from  dark-photon and other experiments. 
For example, the bound on the coupling $\alpha_\mu$ from the  $K^+\to \pi^+ + missing~energy$ decay is at the level $\alpha_{\mu} \leq O(10^{-3})$, which is several orders of magnitude below the value from Eq.(4). 
The visible decay  $Z_{\mu} \rightarrow e^+e^-$  can also occur  at the one-loop 
level.  Its branching fraction is  estimated to be 
Br$(Z_{\mu} \rightarrow e^+e^-) = O((\frac{\alpha}{\pi})^2\alpha_\mu) 
= O(10^{-5})\alpha_\mu$. As a consequence,  in any experiment 
using electrons or quarks as a source of $\zm$`s, the number of 
$\zm \to \ee$ signal events is  suppressed by a  factor  
$ O((\frac{\alpha}{\pi})^4)\approx 10^{-10}$,    resulting in a very weak 
 bound on $\alpha_{\mu}$ (here, the factor $(\frac{\alpha}{\pi})^2$ cames from 
the $\zm$ production).
Finally, we note that if the  $Z_{\mu}$ couples to light dark matter, then an
additional contribution from the invisible  decay mode 
 $Z_{\mu} \to dark~ matter$ that   
increases the $Z_{\mu} \rightarrow  invisible$ decay rate is possible. 
Such a scenario requires additional study, which is beyond the scope of this work.   

To conclude this section, let us stress that existing experimental data restrict  the 
explanation of the $g_{\mu} - 2$ anomaly due to existence of new light gauge boson rather strongly,  
but they do not completely eliminate it. For the model with the interaction \eqref{intmu} the realization 
with invisible $Z_{\mu}$-boson decays into new light $\chi$ --particles for $M_{Z_{\mu}} = 30~ - ~50$ MeV and around 
$M_{Z_{\mu}} = 140$ MeV is still possible. Moreover for the interaction of the $Z_{\mu}$ boson with $L_{\mu} - L_{\tau}$ 
current bounds are rather weak,  and a light $Z_{\mu}$- boson with a mass $M_{Z_{\mu}} \leq 400$ MeV is not excluded as  the source of the $g_{\mu} - 2$ discrepancy.  
\begin{figure}
\includegraphics[width=0.5\textwidth]{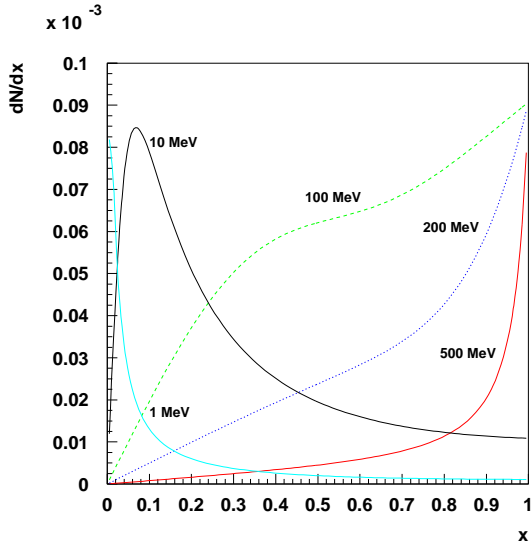}
\caption{Calculated  distributions of the $Z_\mu$ fractional energy $x=E_{Z_\mu}/E_\mu$ from 
the reaction $\react$ at a muon beam energy $E_{\mu}=150$ GeV  for different $Z_\mu$ masses indicated 
near the  curves. The spectra are normalized to a common maximum.}
\label{zspectrum}
\end{figure}

\section{The $\zm$ production and decays}
As the source of $\zm$s, we use bremsstrahlung $\zm$`s produced in the reaction 
\begin{equation}
\mu(p) ~+~ Z(P) \rightarrow Z(P^{'}) ~+~\mu(p^{'})~  + ~Z_{\mu}(k)
\label{react}
\end{equation}
of high-energy muons scattering off the nuclei of a target, as  shown in  Fig. \ref{diagr}.
Here $p,P,P^{'},p^{'},k$ are the four-momenta of incoming muon, incoming $Z$ nuclei, 
outgoing  $Z$ nuclei, 
outgoing  muon and outgoing $Z_{\mu}$ boson, respectively.
\begin{figure}
\includegraphics[width=0.5\textwidth]{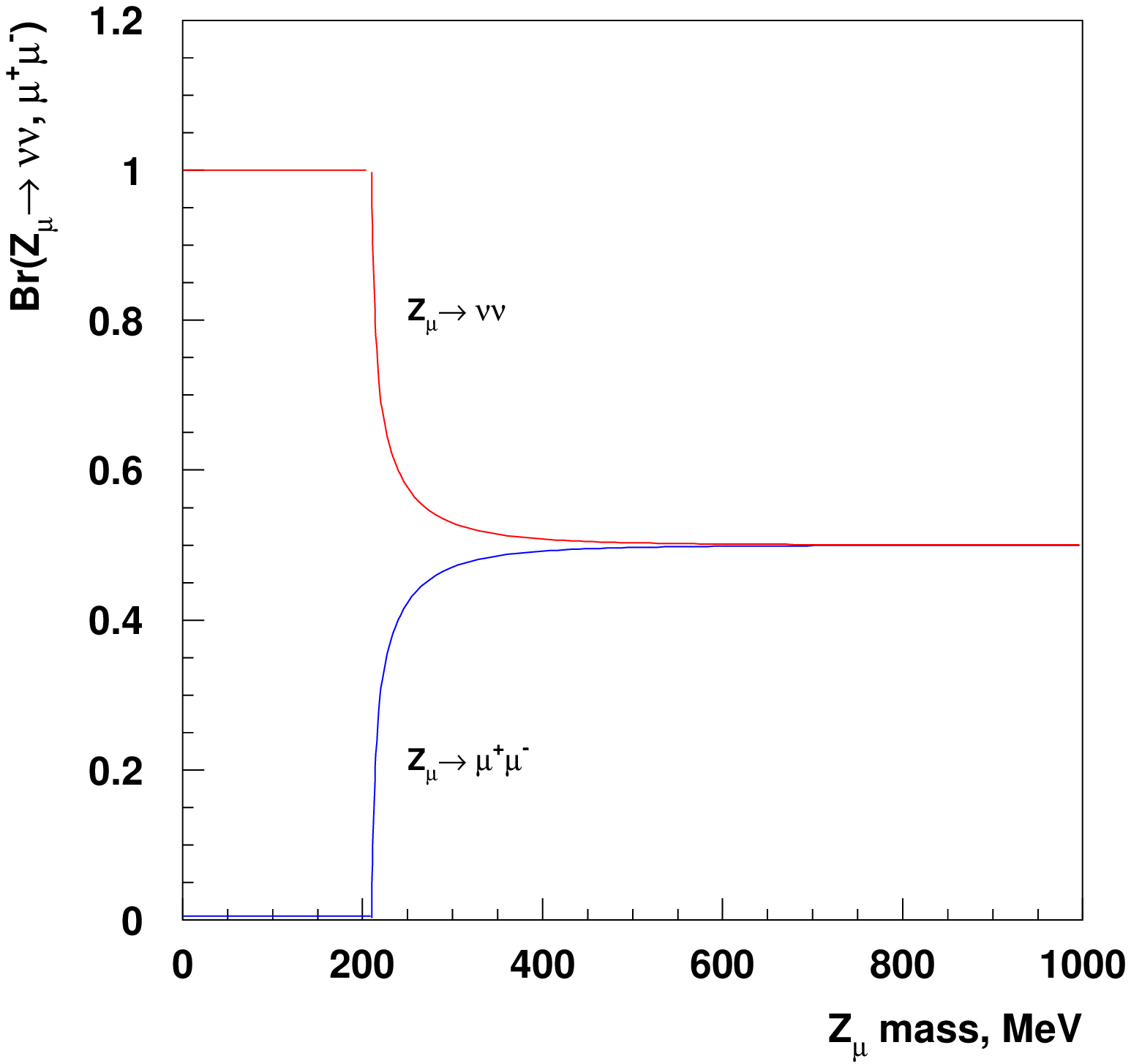}
\caption{The branching fraction of the decays Br$(\zmnn)$ and Br$(\zmvis)$ as a function of the $\zm$ mass.}
\label{branch}
\end{figure}
In this section we give the main formulas for the production of the $Z_{\mu}$ boson in the reaction of Eq.(\ref{react}).
In the Weizsacker-Williams approximation 
\cite{WW}, in the rest frame of the nuclei ($P = (M, 0)$, 
$p = (E_0, \vec{p})$, $p^{'} = (E^{'}, \vec{p}^{'})$, and $k = (E_{Z_{\mu}}, \vec{k})$,  
the $Z_{\mu}$-production cross section at the nuclei 
\begin{equation}
\frac{d\sigma(\mu  + Z \rightarrow \mu + Z_{\mu} +Z)}{dE_{Z_{\mu}}d\cos\theta_{Z_{\mu}}}
\end{equation}
is related to the cross section for real photon scattering, 
$\mu(p)  \gamma(q) \rightarrow \mu(p^{'}) Z_{\mu}(k)$ with $q = P^{'} - P$;
namely,  the following formula allpies:
\begin{eqnarray}
\frac{d\sigma(\mu  + Z \rightarrow \mu + Z_{\mu} +Z)}{dE_{Z_{\mu}}d\cos\theta_{Z_{\mu}}} = 
 \frac{\alpha \chi}{\pi}\frac{E_0x\beta_{Z_{\mu}}}{1-x} \times \\ \nonumber 
\frac{d \sigma (p + q \rightarrow p^{'} +k)}{d(pk)}|_{t = t_{min}}
\end{eqnarray}
Here 
\begin{equation}
x \equiv  E_{Z_{\mu}}/E_0 \,,
\end{equation}
\begin{equation}
t \equiv - q^2 \,,
\end{equation}
\begin{equation}
\beta_{Z_{\mu}} = \sqrt{(1 - m^2_{Z_{\mu}}/E^2_0)} \,
\end{equation}
and $\chi$ is the  effective flux of photons integrated from $t =t_{min}$ to $t_{max}$  \cite{WW}.
The kinematics is determined at $t = t_{min}$. For a given $Z_{\mu}$ momentum the virtuality 
$t$ has its minimum value $t_{min}$ when $\vec{k}$ is collinear with the three-vector $\vec{k} - 
\vec{p}$  \cite{BJORKEN}. One can find \cite{BJORKEN} that 
\begin{equation}
q^0 = \frac{|\vec{q}|^2}{2M} \approx 0 \,,
\end{equation}
\begin{equation}
|\vec{q}| = \frac{U}{2E_0(1-x)} \,,
\end{equation}
where
\begin{equation}
U \equiv U(x, \theta_{Z_{\mu}})=  E^2_0 \theta^2_{Z_{\mu}} x + m^2_{Z_{\mu}}\frac{1-x}{x} +m^2_{\mu}x
\end{equation} 
The Mandelstam variables at $t = t_{min}$ have the form \cite{BJORKEN}
\begin{equation}
-\tilde{u} = m^2_{\mu} - u_2 = 2p \cdot k - m^2_{Z_{\mu}} = U \,,
\end{equation}
\begin{equation}
\tilde{s} = -m^2_{\mu} + s_2 = 2p^{'} \cdot k + m^2_{Z_{\mu}} = \frac{U}{1 - x} \,,
\end{equation}
\begin{equation}
t_2  = (p -p^{'})^2 = -\frac{Ux}{1 - x} + m^2_{Z_{\mu}} \,.
\end{equation} 
\begin{figure*}
\includegraphics[width=0.95\textwidth]{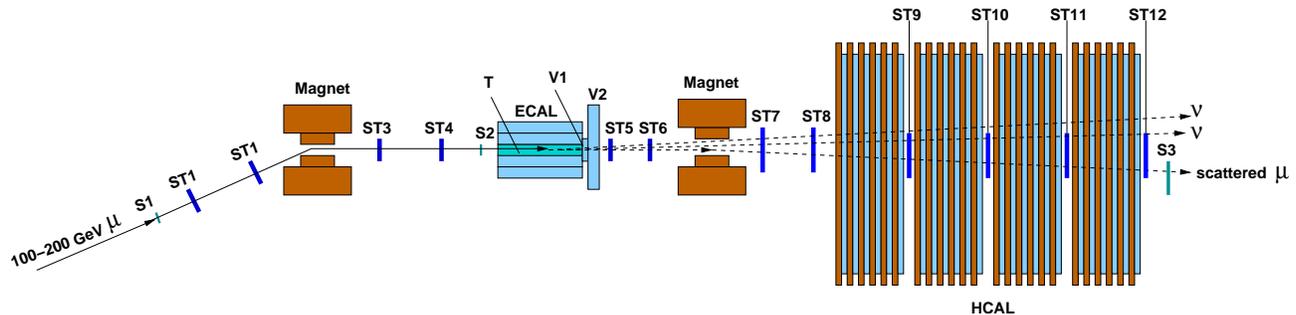}
\caption{ Schematic illustration of the setup to search for dark $\zm$. 
The  bremsstrahlung $\zm$s are produced in the froward direction in the reaction  $\react$ of a 
high-energy muon scattering off nuclei  of an active target $T$. The target is surrounded by an ECAL serving as a veto against photons or other secondaries emitted at a large angle.  
 A fraction $f\lesssim 0.7$, of the   primary beam energy is carried away  by the scattered muon,  while  the  rest of the total energy is  transmitted by the $\zm$ decay neutrino  through the $T$, the veto counters V1 and V2, and a massive hermetic HCAL. The neutrino from the $\zmnn$ decay penetrates them without interactions resulting in a zero-energy   signature in the detector. The dashed line represents the trajectory of 
 the outgoing  muon which passes through the central HCAL cell without interactions. The momentum  of the incident muon is measured by a magnetic spectrometer, while the momentum of the scattered muon is measured by the second one, located downstream of the ECAL (see text).}
 \label{setup}
\end{figure*}
For the  case of a muon beam (contrary to  Ref.\cite{BJORKEN} where an electron beam was studied) 
we cannot neglect the muon mass compared to the $Z_{\mu}$ mass and the $2 \rightarrow 2 $ differential cross section has the form 
\begin{eqnarray}
\frac{d\sigma}{dt_{2}} = \frac{2\pi \alpha \alpha_\mu}{\tilde{s}^2} 
[\frac{\tilde{s}}{-\tilde{u}} + \frac{-\tilde{u}}{\tilde{s}}  
+ 4(\frac{m^2_{{\mu}}}{\tilde{s}} + \frac{m^2_{{\mu}}}{\tilde{u}})^2
&+ 4(\frac{m^2_{{\mu}}}{\tilde{s}} + \frac{m^2_{{\mu}}}{\tilde{u}}) \nonumber \\
+\frac{2m^2_{Z_{\mu}}t_2}{-\tilde{u}\tilde{s}} + 
2m^2_{Z_{\mu}}m^2_{\mu}((\frac{1}{\tilde{s}})^2 + (\frac{1}{\tilde{u}})^2)].&
\end{eqnarray}
In the Weizsacker-Williams approximation the cross section of the 
$\mu(p) ~+~ Z(P) \rightarrow Z(P^{'}) ~+~\mu(p^{'})~  + ~Z_{\mu}(k)$ 
reaction is given by 
\begin{equation}
\frac{1}{E^2_0x}\frac{d\sigma}{dxd\cos\theta_{Z_{\mu}}} =
     4(\frac{\alpha^2\alpha_\mu\chi\beta_{Z_{\mu}}}{1-x})[\frac{C_2}{U^2}
+ \frac{C_3}{U^3} + \frac{C_4}{U^4}] \,,
\end{equation}
where 
\begin{equation}
C_2 =  (1-x) + (1-x)^3 \,,
\end{equation}
\begin{equation}
C_3 =  -2x(1-x)^2 m^2_{Z_{\mu}} - 4m^2_{\mu}x(1-x)^2 \,,
\end{equation}
\begin{equation}
C_4 =  2m^4_{Z_{\mu}}(1-x)^3  + (1-x)^2[4 m^4_{\mu}x^2 + 2 m^2_{\mu}m^2_{Z_{\mu}}(x^2 + (1-x)^2)]\,.
\end{equation}
By  integrating with respect to $\theta_{Z_{\mu}}$,  we find that
\begin{equation} 
\frac{d\sigma}{dx} = 2(\frac{\alpha^2\alpha_\mu\chi\beta_{Z_{\mu}}}{1-x})
[\frac{C_2}{V} +  \frac{C_3}{2V^2} + \frac{C_4}{3V^3}] \,,
\end{equation}
where
\begin{equation}
V =  U(x, \theta_{Z_{\mu}}= 0) = m^2_{Z_{\mu}}\frac{1-x}{x} +m^2_{\mu}x
\end{equation} 
For a general electric form factor $G_2(t)$ \cite{WW}, the effective flux of photons $\chi$ is
\begin{equation}
\chi   = \int^{t_{max}}_{t_{min}} dt \frac{(t - t_{min})}{t^2}G_2(t) \,.
\end{equation}
Note that for heavy atomic nuclei $A$ we also have to take into account the inelastic nuclear 
form factor. Numerically, $\chi = Z^2 \cdot Log$, where the function 
$Log$ depends weakly on atomic screening, 
nuclear size effects  and kinematics  \cite{BJORKEN}.
Numerically, $Log \approx (5 - 10) $ 
for $m_{Z_{\mu}} \leq 500$ MeV\cite{WW, BJORKEN}.  
One can see that compared to the photon bremsstrahlung rate, the $\zm$ production rate is suppressed by a factor 
$\simeq \alpha_\mu m_{\mu}^2/\alpha M_{\zm}^2$. 

In Fig. \ref{zspectrum} an example of the expected  distributions of the energy of a $Z_\mu$ produced by a 150 GeV muon 
impinging on the Pb target is shown for different $\zm$ masses.  
  The spectra are calculated  for the coupling $\alpha_\mu = \alpha$. One can see, that for masses $M_{\zm} \gtrsim 100$ MeV the $\zm$ bremsstrahlung distribution is peaked at the  maximal beam energy.

For $M_{z_{\mu}} < 2m_{\mu}$ the decays $Z_{\mu} \rightarrow \mu \bar{\mu}$ are prohibited and the
$Z_{\mu}$ decays mainly into $Z_{\mu} \rightarrow \nu_{\mu}\bar{\nu}_{\mu}, \nu_{\tau}\bar{\nu}_{\tau}$. 
For $2m_{\mu} < M_{Z_{\mu}} < 2m_{\tau}$, in addition to  decays into neutrino pairs 
$Z_{\mu}$ also  decays into $\mu^+\mu^-$ pairs with the decay width
\begin{equation}
\Gamma(Z_{\mu} \rightarrow \mu^- \mu^+) = \frac{\alpha_\mu M_{\zm}}{3}(1 + \frac{2m^2_{\mu}}{M^2_{Z_{\mu}}})
\sqrt{1 -4 \frac{m^2_{\mu}}{M^2_{Z_{\mu}}}} 
\end{equation}
The branching ratio into $\mu^-\mu^+$ pairs is determined by the formula
\begin{equation}
Br(Z_{\mu} \rightarrow \mu^- \mu^+) =  \frac{K(\frac{m_{\mu}}{M_{Z_{\mu}}})}{1 + K(\frac{m_{\mu}}{M_{Z_{\mu}}})} \,,
\label{bratio}
\end{equation}
where
\begin{equation}
K(\frac{m_{\mu}}{M_{Z_{\mu}}}) =  (1 + \frac{2m^2_{\mu}}{M^2_{Z_{\mu}}})\cdot \sqrt{1 -4 \frac{m^2_{\mu}}{M^2_{Z_{\mu}}}}\,.
\end{equation}

 For the coupling of Eqs.(4) and (5), the $\zm$ with the mass $M_{\zm}\gtrsim 100$ MeV is a short-lived particle with the lifetime  $\tau_{\zm} \lesssim 10^{-15}$ s.  In Fig. \ref{branch} the branching fraction of the decays $\zm\to \nu \nu$ and $\zmvis$ are shown as functions of the $\zm$ mass. One can see that for $M_{\zm} \gtrsim 2 m_\mu$, 50\% of the $\zm$`s
decay invisibly into a couple of neutrinos,  while another 50\% decay  into a $\mu^+\mu^-$ pair.
The latter would   result in the muon trident  signature in the detector. For $\zm$ energies $E_{\zm}\simeq 100$ GeV,  the opening angle  $\Theta_{\mm} \simeq M_{\zm}/E_{\zm}$ of the decay $\mm$ pair is still big enough and the decay muons could be resolved in two separated tracks, so the pairs would be  mostly  detected as double-track events. 
However, the main problem for  the search for the $\zmvis$ decay is the background of muon trident events 
from the QED reaction $\mu Z \to \mu Z \mm$, whose rate substantially exceeds the rate of the reaction \eqref{react}. An additional study, which is beyond the scope of this work, is required for this decay channel. Here, we mostly focus on the case when the reaction \eqref{react} is accompanied by the decay $\zm \to \nu \nu$, resulting in the invisible final state. 

\section{The experiment to search for the $\mu + Z \to \mu + Z + \zm, \zm \to \nu \nu$} \label{sec:ExpInvisible}

The reaction  of the $\zm$ production is a rare  event. For the previously mentioned parameter space, it is  expected to occur with the rate $\lesssim  \alpha_\mu/\alpha \sim 10^{-6}$ with respect to the  ordinary photon production rate. Hence, its observation presents a challenge for the detector design and performance. 

The  experimental setup  specifically designed to search for the $\zm$ production and subsequent decay $\zmnn$ from the reaction of Eq. (\ref{react}) of high-energy muon scattering off  nuclei in a high density target $T$ is schematically shown in Fig. \ref{setup}.  The experiment could employ the upgraded muon beam at the CERN SPS described in details in 
 Ref.\cite{lau}. The  beam was designed to transport high fluxes of muons of the maximum momenta 
in the range between 100 and 225 GeV/c that could be derived from a primary proton beam of 450 GeV/c
with the intensity between 10$^{12}$ and 10$^{13}$ protons per SPS spill.   
The beam  is produced by protons impinging  on a primary beryllium target and transported to the detector in an evacuated beam-line tuned to a freely adjustable  beam momentum  \cite{sps}. 
The typical maximal intensity for a beam energy $\simeq$ 100 GeV, is of the order of $ 5\times 10^7~\mu^-$ for the SPS spill with $10^{12}$ protons on target. The typical SPS cycle for fixed-target (FT)  operation lasts 14.8 s, including 4.8 s  spill duration. The maximal number of FT cycles is four per minute.  
The hadron contamination in the muon beam is remarkably negligible (below $\pi/\mu \lesssim 10^{-6}$) and the size of the beam at the detector position  is of the order of a few cm$^2$.

The detector shown in Fig. \ref{setup} utilizes  two, upstream and downstream,  magnetic spectrometers (MS)
consisting of  dipole magnets and a low-material budget tracker, which is a set of  straw-tubes chambers, ST1-ST4 and ST5-ST8,  allowing for the  reconstruction and  precise measurements of momenta for incident and scattered muons,  respectively. It also uses scintillating fiber hodoscopes:  S1 and S2 define the primary muon beam, while S3 defines the scattered muons, with  the active target $T$  surrounded by a  high-efficiency  electromagnetic calorimeter (ECAL) serving  as a veto  against photons and other secondaries emitted from the target  at large angles. Downstream of the target the detector
 is equipped with  high-efficiency forward veto counters V1 and V2 with  small central holes, and a massive, completely hermetic hadronic calorimeter (HCAL) located at the end of the setup.  The HCAL has four modules, each with  lateral and longitudinal  segmentation. The central part of the first (last)  module is a cell with the lateral size $\simeq 100  \times 100$ mm$^2$ ( $\simeq 400  \times 400$ mm$^2$), used to detect scattered muons and secondaries emitted in the very forward direction. 
 It is also used for the final-state muon identification. The rest of each
HCAL  module serves  as a dump to completely absorb  and detect the energy  of secondary particles produced in the muon  interactions $\mu^- A \to anything$ in the target. The size of the central cells, straw-tube chambers ST9-ST12 and the counter S3 is determined 
  by the requirement to keep the acceptance for deflected scattered muons with 
momentum in the range 15-100 GeV $\gtrsim 90 \%$. 
For example, the lateral size of the S3 counter should be at least 
$50\times 50$ cm$^2$ and is determined mostly by the deflection angle in  the second magnet and  multiple scattering in the HCAL modules of scattered muons. 
    
 In order to suppress  background due to the detection inefficiency,  the HCAL must be longitudinally completely hermetic. To enhance its hermeticity, the HCAL thickness is chosen to be $\simeq 30 ~\lambda_{int}$ (nuclear interaction lengths).  For searches at low  energies,  Cherenkov counters  to enhance the incoming muon tagging efficiency can be used.
   
The method of the search is as follows.
The bremsstrahlung $\zm$s are produced in the reaction \eqref{react} which occurrs uniformly over the length of the target. A fraction ($f$) of the primary beam energy  $E'_\mu = f E_\mu$  is carried away by the scattered muon which is detected by the second magnetic spectrometer, as  shown in Fig. \ref{setup}, tuned for the scattered muon momentum 
$p'_\mu \lesssim f p_\mu$.  The remaining part of the primary muon energy $(1-f)E_\mu$ is transmitted through the HCAL by the
neutrino from the prompt $\zmnn$ decay  resulting in a zero-energy deposition signal in the detector, i.e. in missing energy $E_{miss}= E_\mu - E'_\mu$. 

The occurrence of $\zm$  produced in $\mu^- Z $ interactions would appear as an excess of 
events with a single scattered  muon accompanied by zero-energy deposition  in the detector,  as shown in 
Fig. \ref{setup},  above those expected from the background sources. The signal candidate events have the signature: 
\begin{equation}
S_{\zm} = {\rm S1 \cdot S2 \cdot T \cdot \mu_{out} \cdot\overline{V1\cdot V2\cdot HCAL}}
\label{signzm} 
\end{equation}
and should satisfy the following selection criteria:  
\begin{enumerate}[(i)]
\item ${\rm S1\cdot S2 \cdot T}$: The presence of an incoming muon with  
energy  150 GeV. The energy deposited in the target is consisted with that expected from the minimum
ionizing particle (MIP).

\item $\mu_{out}$: The presence of a single scattered muon with energy $E'_\mu \lesssim 100$ GeV after the target, and the presence of a single muon track in the straw-tube chambers ST9-ST12 traversing the four HCAL modules. 

\item $\rm{\overline{V1\cdot V2 \cdot HCAL}}$:  No energy deposition in the veto counters V1 and V2, no  energy deposition in the central HCAL cells above those expected from the scattered muon, and no energy in the rest of the HCAL modules. 
\end{enumerate}
\begin{figure}
\includegraphics[width=0.55\textwidth]{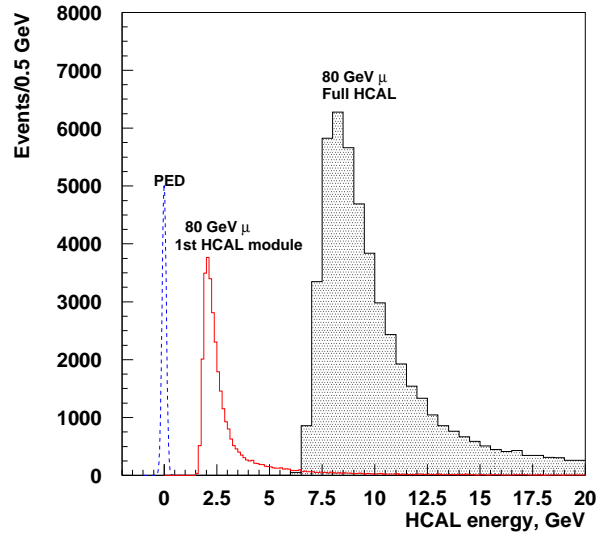}
\caption{The distribution of the energy deposited  in the central cell of the first HCAL module (red histogram) and in all four HCAL central cells (shaded histogram)   by traversing muons with energy $E_{\mu}= 80$ GeV. The peak of the pedestal sum over the rest of the HCAL   in units of hadronic energy is also shown.}
\label{pedmu}
\end{figure}
The "zero-energy" signal is defined by the following 
\begin{enumerate}[(i)]
\item The presence of the energy $E^\mu_{ECAL}$deposited in the target, which is consistent with  that deposited by the MIP. 

\item The presence of energy $E^\mu_{HCAL}$ deposited in the HCAL cell  crossing by the scattered muon compatible with that expected from the MIP, $E^\mu_{HCAL} \simeq E_{mip}$,  see Fig. \ref{pedmu}. The primary muons that do not interact in the target and pass the HCAL without interactions  deposit about (2.5$\pm$1) GeV in each central cell.

\item No energy deposition in the veto counters V1 and V2, and in the rest of the  HCAL, $ E_{HCAL}\lesssim 100$ MeV,  presented by the sum of pedestals of the readout system,  as shown in Fig. \ref{pedmu}. The effective width of the signal in the rest of the HCAL  is $\simeq$ 100 MeV.

\item The total energy deposited in the ECAL and HCAL is $E_{tot}=E_{ECAL}+E_{HCAL}\lesssim 12$ GeV. 
\end{enumerate}

The optimal primary beam energy is selected using the following considerations. First, it has to be   high enough to provide the highest rate for the production of  $\zm$`s in the sub-GeV mass range; second,  it should correspond to as large an absolute value of $E_{miss}$ as possible; and third, it should allow one to operate muon beam at high  intensity. Taking  these  considerations into account, a beam energy around $\simeq$150 GeV is chosen. 

\begin{figure*}
\includegraphics[width=0.8\textwidth]{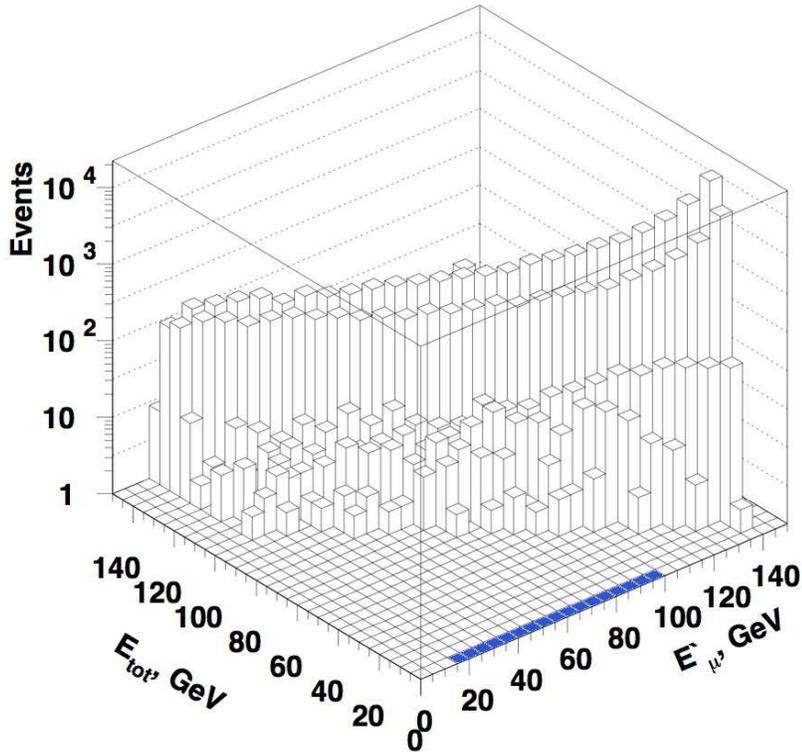}
\caption{Simulated distribution of events in the $(E'_\mu; E_{tot})$ plane from the  bremsstrahlung, knock-on, pair-production and photonuclear muon interactions in the target that passed the veto selection criteria (v).  The primary muon beam energy is $E_{\mu}=150$ GeV and the total number of incident muons is $n_\mu \simeq 3\times 10^8$. The blue  area  shows  the signal of the reaction  $\mu + Z \to \mu + Z + \zm, \zm \to \nu \nu$:
the  scattered muon energy  $15\lesssim E'_\mu \lesssim 100 $ GeV and $E_{tot}=E_{ECAL}+E_{HCAL} \lesssim 12$ GeV.}
\label{hcalbiplot}
\end{figure*}

In Fig. \ref{hcalbiplot}, the simulated distribution of the events from the  bremsstrahlung, knock-on, pair-production and photonuclear muon interactions in the target in the  $(E'_\mu; E_{tot})$ plane is shown for the primary muon beam energy $E_{\mu}=150$ GeV and a total number of incident muons $n_\mu \simeq 3\times 10^8$ (see Sec.V). The events are selected with the requirement of no energy deposition in the veto counters V1 and V2. The signal of the reaction  $\mu + Z \to \mu + Z + \zm, \zm \to \nu \nu$ is defined by a  scattered muon energy of $E'_\mu \lesssim 100 $ GeV and a total energy $E_{tot}=E_{ECAL}+E_{HCAL} \lesssim 12$ GeV.  The width of the  signal region along the $E_{tot}$ axis corresponds to an energy deposition around 2.4 GeV in each 
 central cell of four consecutive HCAL modules, as illustrated  in Fig. \ref{pedmu}, plus about 0.5 GeV deposited in the ECAL, i.e. around 10 GeV for the total energy. 
\section{Background}
To estimate the background and  sensitivity of the proposed experiment, 
 a simplified feasibility study  based on GEANT4 \cite{geant}
Monte Carlo simulations has been  performed for incoming muons with an energy of 150 GeV. In these simulations 
the target is the radiation-hard shashlik  module  ($X_0 \simeq 1.5$ cm) with a total thickness of about 50 X$_0$,
surrounded by  the ECAL, which  is  a hodoscope  array of  the lead-scintillator  counters that are also of the shashlik type,   each with a size of $38\times 38 \times 400$ mm$^3$, allowing for accurate measurements of the lateral energy leak from the target. The shashlik calorimeter is a  sampling calorimeter
in which scintillation light is read out though  wavelength-shifting  fibers running perpendicular to the absorber plate; see, e.g. Ref. \cite{shashlyk}. 
 The target module consists of 300 layers of  1 mm thick lead and 1 mm thick plastic scintillator plates and has  longitudinal segmentation.

Each  of the scintillator counters S1, S2, and S3  consists of two layers of scintillating fiber strips,  arranged  in the X and Y directions, respectively. Each strip consists of about 100 fibers of 1 mm square. The number of photoelectrons  produced by a MIP crossing the strip is $\simeq$ 
20 photoelectrons.
 The veto counters are assumed to be  1-2 cm thick, high-sensitivity scintillator  arrays with a high light yield  of 
 $\gtrsim 10^2$ photoelectrons per 1 MeV of deposited energy. It is also assumed that the veto inefficiency  for the MIP  detection  is, conservatively, $\lesssim 10^{-4}$. The hadronic calorimeter is a set of four modules. Each module is a sandwich of alternating layers of iron and scintillator with a thickness of 25 mm and 4 mm,  respectively, and with a lateral size $120\times 120$ cm$^2$.    Each module consists of 48 such layers and has a total thickness of $\simeq 7\lambda_{int}$. 
The number of photoelectrons produced by a MIP crossing the module  is in the range $\simeq$ 150-200 photoelectrons.
The energy resolution of the HCAL calorimeters as a function of the beam energy is taken to be 
$\frac{\sigma}{E} \simeq \frac{ 60 \%}{\sqrt{E}}$ \cite{ihepHCAL}. The energy threshold for  zero energy in the HCAL is $\simeq 0.1$ GeV. We assume that 
the momenta of the in- and outgoing muons are measured with a precision of a few percent. 
 The scattered muon produced in the target is defined as a single track crossing the  HCAL and the straw-tube stations ST9-ST12 and  accompanied by no activity in the HCAL modules.
\begin{figure*}
\includegraphics[width=0.5\textwidth]{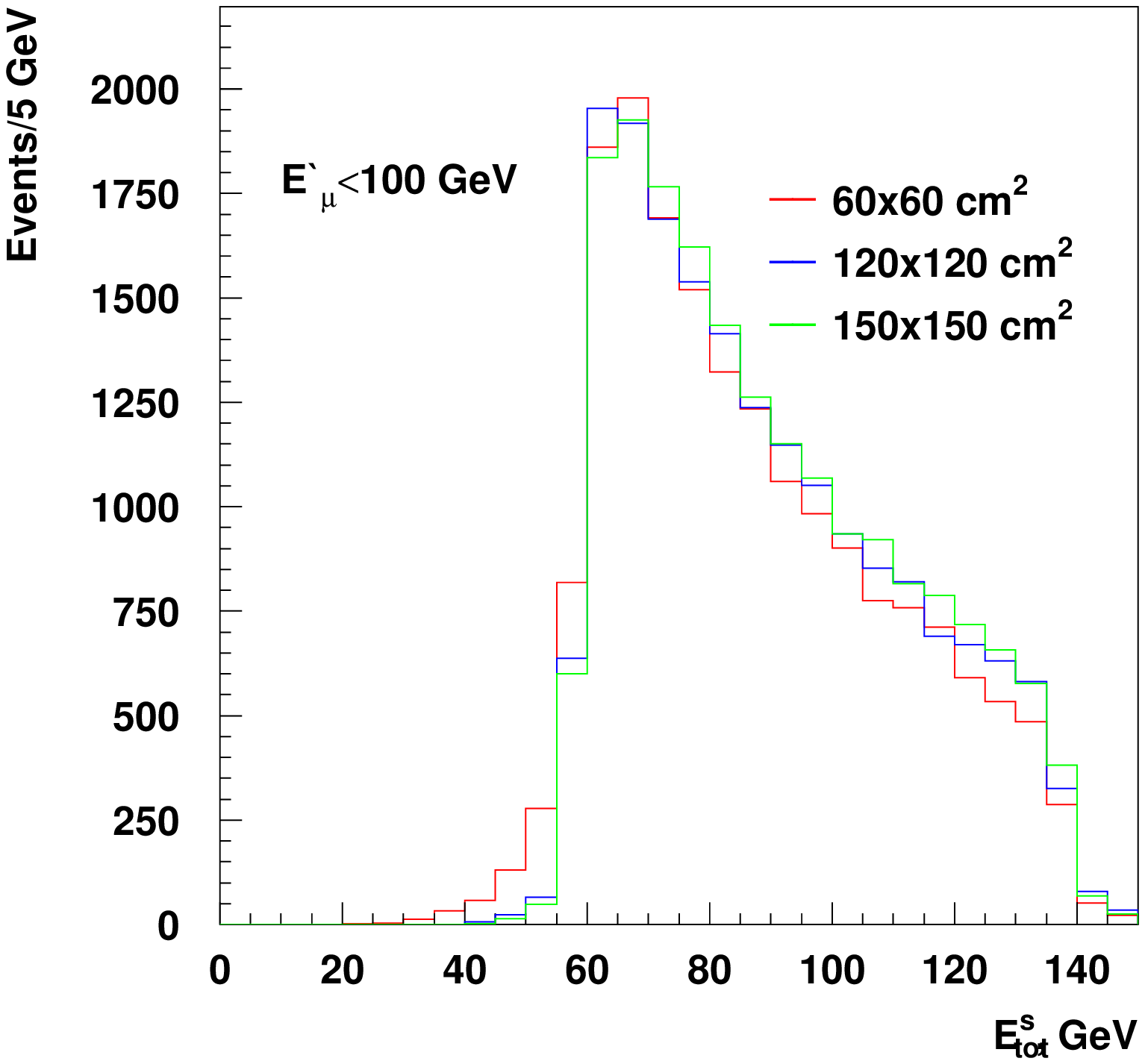}
\includegraphics[width=0.5\textwidth]{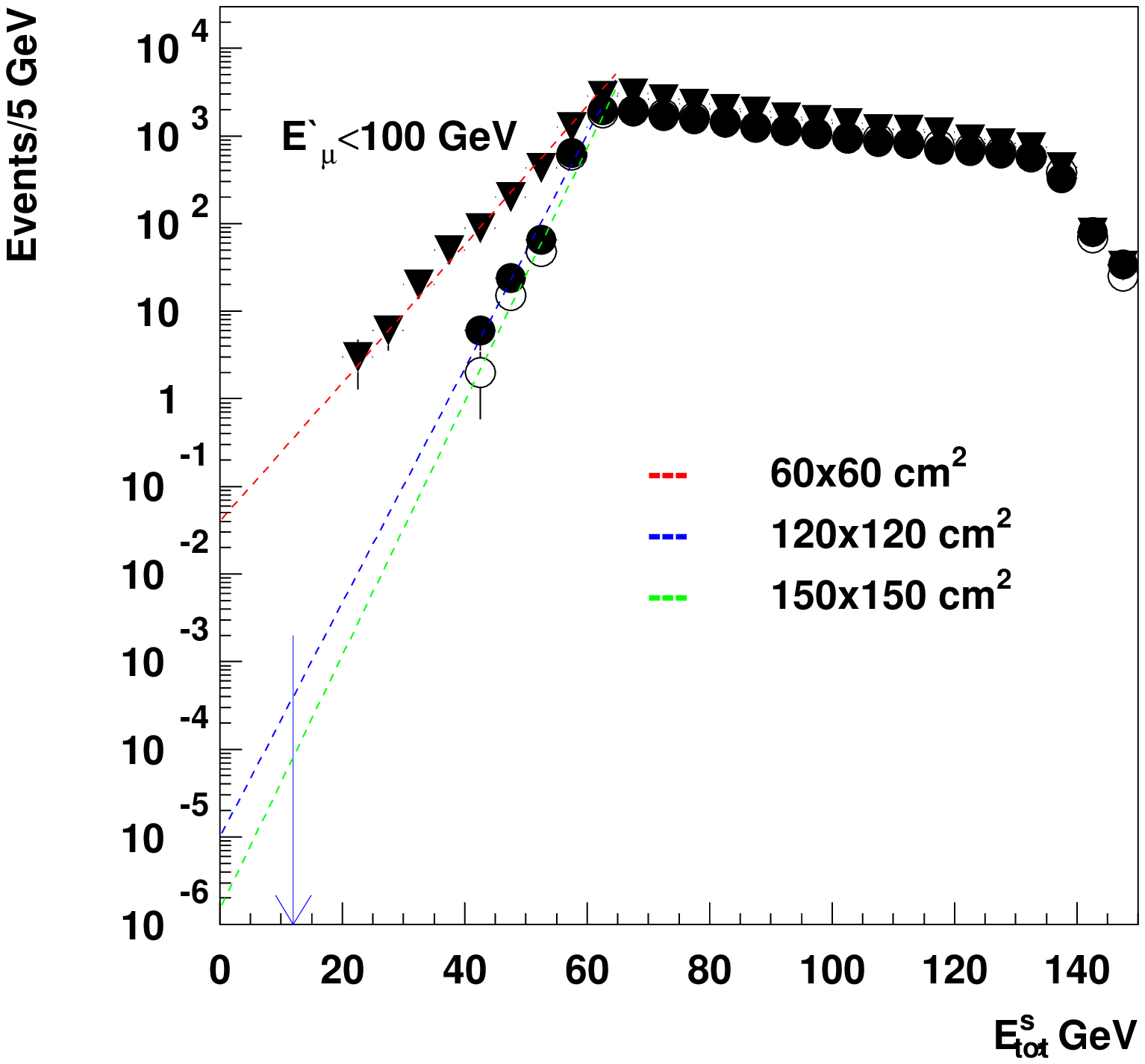}
\caption{ The l.h.s. shows the simulated distribution of the energy deposited  in the ECAL+HCAL 
by secondaries  from the  bremsstrahlung,  pair-production, knock-on and photonuclear muon interactions in the target.  The events are selected by requiring the presence of a scattered muon crossing the central HCAL cells with initial energy  $E'_\mu \lesssim 100 $ GeV and no energy deposition in the veto counters V1 and V2.
  The energy ($E^\mu_{ECAL}+E^\mu_{HCAL}$) deposited by the scattered muon in the ECAL and the HCAL central cells is subtracted.   On the  r.h.s. the same distribution (dots) is shown on a  logarithmic scale.  Other plots correspond to the energy distribution in the ECAL+HCAL for the HCAL with half of the  lateral size, i.e., 
$60\times 60$ cm$^2$ (triangles) and $150\times 150$ cm$^2$ (open circles). The curves are the fit of the  low-energy tail of the distributions by a smooth polynomial function  extrapolated to the signal  region  $E^{s}_{tot}=E_{ECAL}+E_{HCAL}-E^\mu_{ECAL}-E^\mu_{HCAL}\lesssim 12$ GeV, indicated by the arrow, in order to  conservatively evaluate  the expected number of background events. The importance of the HCAL transverse size for the minimization of the  
lateral leak of the energy and the reduction of  the number of background events is clearly seen.}
\label{hcalextra}
\end{figure*}   
The background reactions resulting in  the signature of Eq. (\ref{signzm}) can be classified as being due to physical-  and  beam-related sources.  To investigate these backgrounds down to the level  $ \lesssim 10^{-10}$ with the full detector simulation  would require  a prohibitively large amount of computer time. Consequently, only the following background sources - identified as the most dangerous- are considered and evaluated  with  reasonable statistics combined  with numerical calculations:

\begin{itemize}
\item One of the main background sources is related to the  low-energy tail in the energy distribution of beam muons. The muon energy  is lost due to the interaction of the particles with passive material, such as, e.g.,  entrance windows and the  residual gas of  beam lines. Another source of low-energy muons  is due to the   in-flight decays of pions and kaons that contaminate the beam.  The uncertainties arising from the lack of knowledge of the dead material composition in the beam line  are potentially the largest source of systematic uncertainty in accurate calculations  of the fraction  and energy distribution of these events. An estimation shows  that the fraction of events with energy below  $\lesssim 100$ GeV in the muon  beam tuned to 150  GeV could be as large as  $10^{-7}$. Hence, the sensitivity of the experiment could be  determined by the presence of such muons in the beam, unless one takes special  measures to suppress this background.

To improve the high-energy muon selection  and suppress the background from the possible admixture of low-energy muons, an additional  tagging system utilizing  a magnetic spectrometer is used,  as schematically shown in  Fig. \ref{setup}.
The precision of the muon momentum measurement with four straw-tube chambers is 
dominated by the track measurement errors $\sigma (x)(\simeq  100$ mkm) and is given by \cite{grupen}
\begin{equation}
\frac{\sigma(p)}{p}\simeq \frac{\sigma (x)[m]\cdot 8 p [GeV]}{0.3 B[T] (L[m])^2}
\end{equation} 
where $B$  and $L$  are the field strength and length  of the magnet.    The contribution form 
 muon multiple scattering is negligible. Taking into account that $B\simeq$ 2 T, and $L\simeq$ 2 m results in 
$\frac{\sigma (p)}{p} \simeq 3 \%$ for muons with momentum $p=100$ GeV. 
Thus, the probability for a muon with $p\lesssim 100$ GeV to be taken as a 
150 GeV one corresponds to the $\gtrsim 15$ sigma level and is negligible.    
 The overall suppression of this background by the  H4 beam-line spectrometer combined with this additional one is expected to be at a level much  below $10^{-13}$ per incident muon.
 
\item The low-energy  muons  could appear in the  beam after the target  due to the in-flight 
$\pi \to \mu \nu$ decay 
of the punch-through 150 GeV pions  in the region between the  tracker stations ST3 and ST4. In this case the pion could mimic 
the primary muon, while the decay muon could be taken as a fake scattered muon. Taking into account that the admixture of the pion in the beam is at the level 
$P_\pi \lesssim 10^{-6}$ \cite{lau} and  the probability for the pion to decay 
at a distance of 4 m between the two spectrometers 
 $P_d \sim 5\times 10^{-4}$ results in an overall expected background at the level of $5\times 10^{-10}$ per incoming muon.
 To suppress this background further, one can use a cut that requires the maximal scattered muon energy to be below the minimal kinematically allowed  decay muon energy  $E_{min}^\mu \simeq $ 86 GeV.

However, in order to keep the muon efficiency as high as possible and still use 
the 15-100 GeV signal window shown in Fig.\ref{hcalbiplot}, one can 
reduce the hadron contamination in the muon beam by utilizing  an additional 
 hadron absorber installed in the upstream part of the beam line. 
Using of a $\sim$ 300 cm thick beryllium filter,  which has the optimal 
ratio of $\lambda_{int}(\simeq$40 cm)$/X_0(\simeq$ 35~cm),  results in an 
additional reduction of the $\pi/\mu$ ratio down to $P_\pi \lesssim  10^{-9}$, 
at the cost of  a small  muon flux attenuation and an increase of the 
 average multiple scattering angle. The combined probability for the $\pi \to \mu \nu$ decay background  is then  $P_\pi P_{dec} P_{cut} \lesssim 10^{-13}$, 
where  $P_{cut}\lesssim 0.2$ is a probability for decay muon to have 
momentum $P_\mu \lesssim 100 $ GeV. 

\item
The fake signature~ of Eq.\eqref{signzm} could also arise when  a high-energy muon loses energy  through 
hard bremsstrahlung (BR), knock-on electrons (KN),  pair-production (PP)  or photonuclear (PN) interactions in the target.The fraction of these reactions, compared with the total muon energy losses including ionization losses,  depends on the ratio $E'_\mu/E_\mu$ and for the Pb target is in the range $\simeq 10^{-3}-10^{-5}$ per $X_0$ for
 $0.1 \lesssim E'_\mu/E_\mu\lesssim 0.9$ \cite{moh}.   Such reactions
could yield a low-energy scattered muon accompanied by   neutral penetrating particles in the final state (e.g. photons,  neutrons, $K^0_L$, etc.),  which then could escape detection in the rest of the detector. Simulations show that in this case, the background is dominated  by the photonuclear reactions accompanied by the  
emission of hadrons or  a leading hadron  $h$  from the muon-induced  reactions $\mu A \rightarrow \mu h X$ which could escape detection due to incomplete hermeticity of the  HCAL. For the energy range discussed the  muon photonuclear cross section is $\sigma_{PN}(\mu N \rightarrow  \mu X)\simeq 10^{-2} \sigma_{tot}$ of the total interaction cross section 
$\sigma_{tot}=\sigma_{BR}+\sigma_{KN}+\sigma_{PP}+\sigma_{PN}$ \cite{moh}.  This important source of background was examined by using several methods.

\item
 In Fig. \ref{hcalextra} we show the  simulated distribution of the energy deposited  in the (ECAl+HCAL)  by secondaries  from the  bremsstrahlung, knock-on, pair-production or photonuclear muon interactions in the target. The events are selected by requiring additionally the presence of a scattered muon crossing the central HCAL cells with an energy  $E'_\mu \lesssim 100 $ GeV and no energy deposition in the veto counters V1 and V2. The energy deposited by the scattered muon in the ECAL and the HCAL central cell is subtracted. The  low-energy tail of this distribution  was fitted by a smooth polynomial function and extrapolated to the energy region $E_{tot}\lesssim 12$ GeV  to evaluate the number of background events in the signal region. Those events with an energy deposition below the typical energy deposited by the MIP
  could mix with the muon signal resulting in the fake signal. Using this rough estimate we find that this background is expected to be at the level $\lesssim 10^{-12}$ per incoming  muon. IN the same plot the distribution of the energy 
  in the ECAL+HCAL for the HCAL with half the lateral size, i.e. 
$60\times 60$ cm$^2$ and $150\times 150$ cm$^2$ are shown for comparison. The effect of the HCAL transverse size on the lateral leak of the energy deposition from the  bremsstrahlung, 
pair-production, knock-on or photonuclear muon interactions in the target and  the corresponding number of background events is clearly seen.

\begin{figure*}
\includegraphics[width=0.5\textwidth]{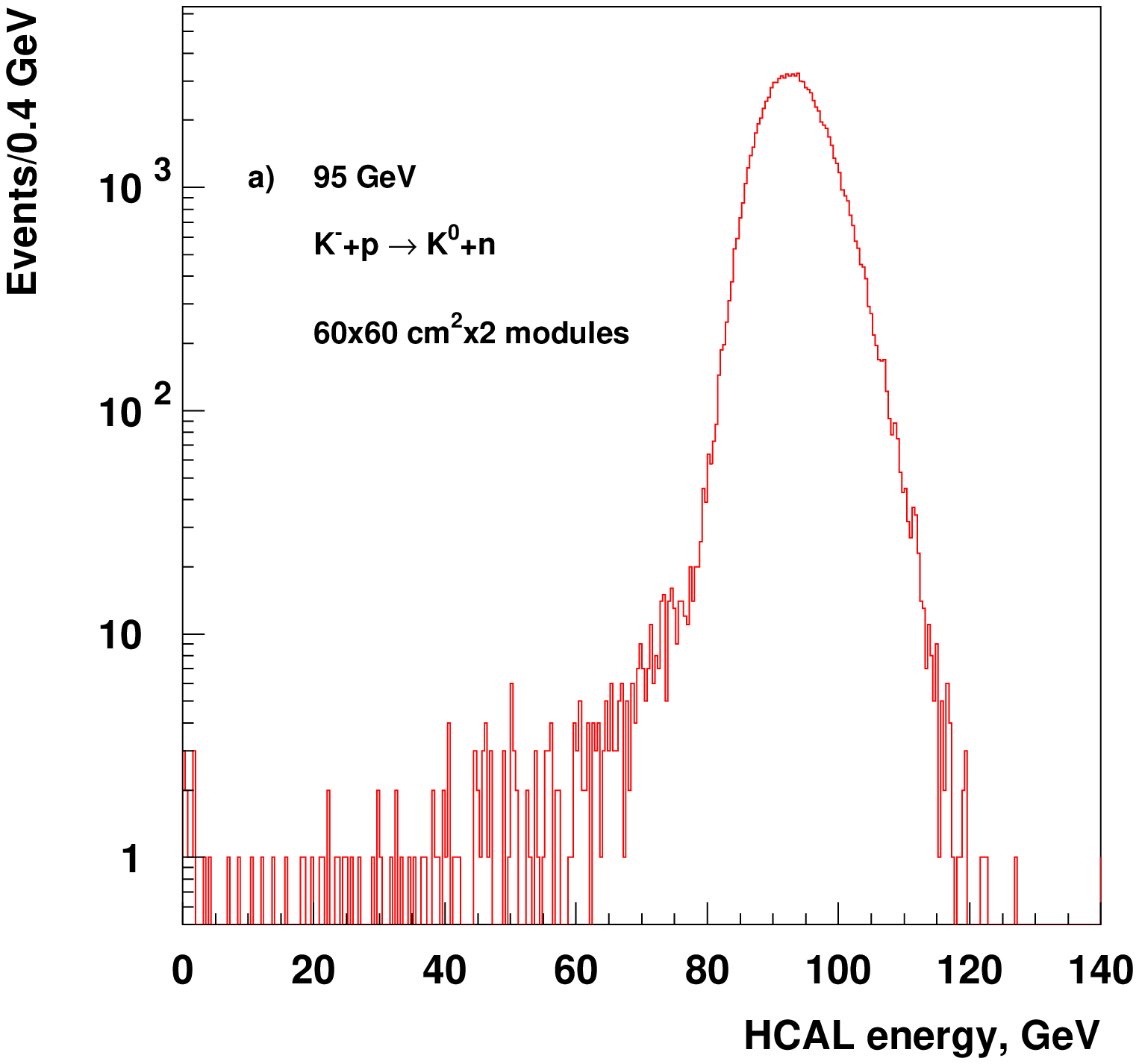}
\includegraphics[width=0.5\textwidth]{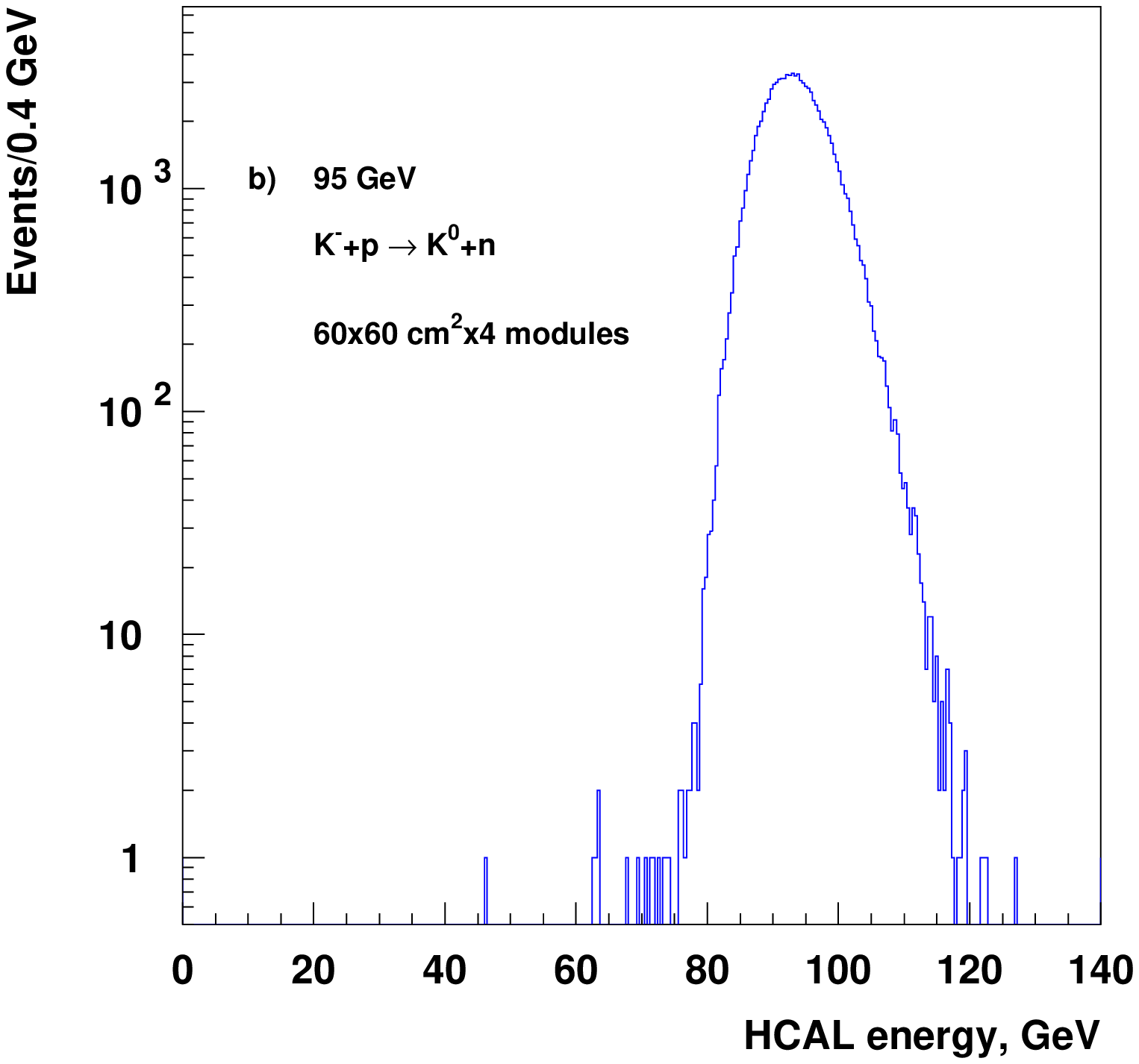}
\caption{  Expected distributions of energy deposited by $\simeq10^6~K^0$ with energy $\simeq$ 95 GeV  in two (a)  and four (b) consecutive HCAL modules. The peak at zero energy in spectrum (a) is due to the punch-through neutral kaons.
}
\label{punch}
\end{figure*}    
\item Another method is based on the direct estimate of the probability for the large missing energy in the detector.   The energy could also leak when the  leading neutron or $K^0_L$ punches through the  HCAL without depositing energy above a certain threshold $E_{th}$. In this case, if the sum of the energy released in the HCAL is below $E_{th}$, the event is considered as a "zero-energy" event.
 The punch-through  probability $P_{pth}$  is defined roughly by $P_{pth}\simeq exp(-L_{tot}/\lambda_{int})$,
where $L_{tot}$ is the HCAL length. As discussed previously, it can be suppressed by using  the HCAL with a thickness of  $\simeq 30 \lambda_{int}$, resulting in a $P_{pth}$ of  $\ll 10^{-10}$.  
This value  should be multiplied by a  conservative factor $\lesssim 10^{-4}$, which is the probability of a  single leading hadron  production in the target, resulting in the final estimate of  $\lesssim 10^{-13}$ for the level of this background per incoming muon. 

\item For completeness, the HCAL nonhermeticity and corresponding background for high-energy secondary hadrons were cross-checked  with  GEANT4-based simulations in the following way. In Fig. \ref{punch} we shoe the expected distributions of energy deposited by $\simeq10^6~K^0$ with energy $\simeq$ 95 GeV   in two (a)  and four (b) consecutive HCAL modules. The peak at zero energy in the spectrum (a) is due to the punch-through neutral kaons, while for the full HCAL thickness there are no such missing energy events in distribution (b).

For another sample of simulated events, the low-energy tail in the  distribution of energy deposited by 
$\simeq 10^7$  150 GeV neutrons in the HCAL was fitted by a smooth polynomial function and extrapolated to the lower-energy region  in order to evaluate the number of events below a certain threshold $E_{th}$. This procedure results in an estimate for the HCAL nonhermeticity, defined as the ratio of the number of events below the threshold $E_{th}$ to the total number of incoming particles: $\eta = n(E<E_{th})/n_{tot}$. 
For the energy threshold $E_{th} \simeq 0.1$ GeV the nonhermeticity is expected to be at the level 
$\eta \lesssim 10^{-9}$.  Taking into account a probability of  producing a  single leading hadron per incoming muon of $P_{h} \lesssim 10^{-4}$, results in an overall level of this background of $\lesssim10^{-13}$, in agreement with the previous rough estimate.

\item
The fake signature of Eq. (\ref{signzm}) could be due to the QED production 
  of muon trident events, $\mu Z \to \mu Z \mm$, with asymmetrical muon momenta in the muon pair. In this case,  the lower-energy  muon could be poorly detected, and another one could admix to the scattered muon in the HCAL central cell. A preliminary simulation study shows 
that this background can be suppressed down to the $10^{-12}$ level, provided the inefficiency of veto counters V1 and V2 is below $10^{-4}$ and the two tracks separated by a distance $\simeq 1$ mm are resolved by the ST5-ST8 and ST9-ST12 trackers.    
\end{itemize}

In Table~\ref{tab:table2} contributions from the all dominant background 
processes are summarized for a primary muon beam energy of 150 GeV. The total background is found to be  at the level $\lesssim  10^{-12}$. The contribution 
of additional subdominant background sources (e.g., such as very asymmetric 
$\mu \to e \nu \nu$ decays accompanied by  low-energy muon production in the 
HCAL by the decay electron, cosmic muons, etc.) is negligible.      
This means that the search accumulated up to $\simeq 10^{12}$ $\mu^-$ events is expected to be background free.  
\begin{table}[tbh!] 
\begin{center}
\caption{Expected contributions to the total level of background from different background sources estimated for a beam energy of 150 GeV (see text for details).}\label{tab:table2}
\vspace{0.15cm}
\begin{tabular}{lr}
\hline
\hline
Source of background& Expected level\\
\hline
$\mu$ low-energy tail& $ \lesssim  10^{-13}$\\
HCAL nonhermeticity & $ \lesssim   10^{-13}$\\
$\mu$ induced photonuclear reactions & $\lesssim 10^{-13}$\\
$\mu$ trident events & $\lesssim 10^{-12}$\\
\hline 
Total   &         $ \lesssim 10^{-12}$\\
\hline
\hline
\end{tabular}
\end{center}
\end{table}
\section{Expected sensitivity}
To estimate the expected sensitivities we used simulations of the process shown 
in Fig. \ref{setup}  to calculate the production rate and energy distributions  of muons produced in the target by taking into account the  normalization of the scattered muon yield from the target taken from the original publications \cite{moh}. 
The calculated fluxes and energy distributions  of scattered muons produced in the target are used to predict the number of signal events in the detector. 
For a given total number of primary muons $n_{\mu}$, the expected number of events from the reaction $\mu + Z \to \mu + Z + \zm, \zm \to \nu \nu$ occurring within the decay length $L$ of the detector is given by 
\begin{eqnarray}
&n_{\zm}= k n_{\mu}Br(\zmnn)\frac{\rho N_{av}}{A} \nonumber \\
&\cdot \int\frac{\sigma(\react)}{dx} d \zeta(M_{\zm}) dx~~~ 
\label{nev}
\end{eqnarray}
 with $d=1$ for $M_{\zm}< 2 m_\mu$, and $d=\Bigl[1-{\rm exp}\Bigl(-\frac{LM_{\zm}}{P_{\zm}\tau_{\zm}}\Bigr)\Bigr]$ for $M_{\zm}>2 m_\mu$.  Here, the coefficient $k$ is a normalization factor that was
tuned to obtain the total  cross sections of meson production,  $P_{\zm}$ and $\tau_{\zm}$ are the  produced $\zm$ momentum and lifetime at rest, respectively, $\zeta(M_{\zm})$ is the overall signal reconstruction efficiency, $\rho$
is the density of the target,$L$ is the decay length in the detector,  and $N_A$ is the Avogadro number. In this estimate we neglect the scattered $\mu$ interactions in the target,  the  momentum of the incoming muons is  $<p_{\mu}>\simeq 150$ GeV,  and the efficiency $\zeta (M_{\zm})$ is in the range $\simeq 0.1-0.5$ for the masses 
1 MeV$\lesssim M_{\zm}\lesssim O(5)$ GeV.

\begin{figure}
\includegraphics[width=0.55\textwidth]{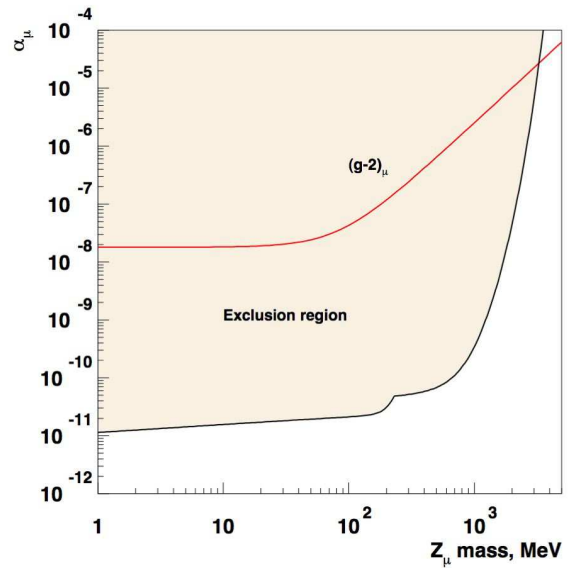}
\caption{Exclusion region in the ($M_{\zm}, \alpha_\mu$) plane expected from the results of the
 proposed experiment for 10$^{12}$ incident muons at the energy $E_{\mu}=150$ GeV.
The red line represents the value of  $\alpha_\mu$ required to explain the muon g-2 discrepancy as a function 
of the $\zm$ mass.}
\label{exclusion}
\end{figure}

The obtained results can be used to impose constraints on the previously discussed coupling strength $\alpha_\mu$
as a function of the $\zm $ mass. 
Using the relation $n_{\zm}^{90\%} > n_{\zm}$, where $n_{\zm}^{90\%}$ (= 2.3 events) is the 90$\%~C.L.$ upper limit for the  number of signal events and Eq. (\ref{nev}), one can  determine 
the expected $90\%~ C.L.$ upper limits from the results of the proposed experiment, which are shown in 
Fig. \ref{exclusion} together with values of the coupling $\alpha_\mu$ required to explain the muon g-2 
anomaly. These bounds are calculated for a scattered muon energy $10 \lesssim E'_\mu \lesssim 100$ GeV and a  total of $10^{12}$ incident muons  in the background-free case. Here we assume an exposure to the muon beam with a nominal rate is a few months.
  
The statistical limit on the sensitivity of the proposed experiment is mostly set  by the number of 
accumulated events. However, there is a limitation factor related to the HCAL signal duration ($\tau_{HCAL} \simeq 100$ ns) resulting in a maximally allowed muon counting rate  of $ \lesssim 1/ \tau_{HCAL} \simeq 10^{6} \mu^- /$s in order to avoid significant loss of the signal efficiency due to the pileup effect.    
To evade this limitation, one could implement a special muon pileup removal algorithm to allow for high-efficiency  reconstruction of the  zero-energy signal properties and the shape in high muon pileup environments,  and run the experiment at the muon beam rate $\simeq 1/\tau_{HCAL}\simeq 10^7~\mu^-/s$.
 Thus,  in the background-free experiment  one could expect a sensitivity in the process $\mu + Z \to \mu + Z + \zm, \zm \to \nu \nu$  that is even higher then those presented above, assuming 
an exposure to the high-intensity muon beam of a few  months. 
 In the case of the $\zm$ signal observation,  several methods could be used to cross-check the result. For instance, 
to test whether the  signal is due to the HCAL nonhermeticity or not, one could perform  measurements with different HCAL thicknesses, i.e., with  one, two, three, and four consecutive HCAL modules.  In this case the  background level  can be evaluated by extrapolating the results to an infinite HCAL thickness. To insure that there is no additional background due to the HCAL transverse hermeticity one could perform  measurements for different distances between 
the target and the HCAL.   
The evaluation of the signal and background  could also be obtained from the results of measurements at different  muon beam energies. 
Finally, we note that the presented analysis gives an illustrative order of magnitude for the sensitivity of the 
proposed experiment and may be strengthened  by more detailed 
simulations of the  experimental setup.\\

\section {Conclusion}

In this work we considered the discrepancy between the measured and predicted values of the 
muon g-2 which could be explained by the existence of a new light gauge boson $\zm$ predominantly coupled to the second and third generations. 
We  proposed  performing an experiment dedicated to the sensitive search for the $\zm$ by using available 
$\simeq 100$ GeV muons beams from the CERN SPS.
If  the $\zm$s  exist, they could be produced in the reaction $\react$  and be observed by looking for events  with a specific  signature, namely those missing a large fraction of the beam energy in the detector.
 A feasibility study of the experimental setup shows that this specific signal of the $\zm$ allows for searches for 
 the $\zm$  with a sensitivity in the coupling constant $\alpha_\mu \gtrsim 10^{-11}$, i.e., 3 orders of magnitude stronger than the value $\alpha_{\mu} \sim 10^{-8} $ explaining  the  3.6$\sigma$ muon $g - 2 $ discrepancy for the $\zm$ mass range $M_{\zm}< O(5)$ GeV \cite{theor5}. 

These results could be obtained with a detector  optimized for  
several of its properties, namely, i) the intensity and purity of the primary pion and kaon beams,  ii) the high efficiency of the 
veto counters, and  iii) the high level 
of hermeticity for the hadronic  calorimeters.
Large amount of high-energy muons and high background suppression are crucial for  improving the sensitivity of the search. To obtain the best limits,  the choice of the energy and intensity  of the beam, as well as the background level should be compromised.  

We point out  
that the availability of high-energy and -intensity muon beams at CERN SPS provides a
unique opportunity to either discover or rule out the $\zm$ with the proposed search in the near future.  
 The experiment is based on the missing-energy approach developed for the searches for
invisible decays of dark photons and (pseudo)scalar mesons at CERN \cite{sngldms,ldms,snginv} and is complementary to these experiments. It also  provides interesting  motivations for further muon  studies and fits well with 
  the present  muon physics program at CERN. 

{\large \bf Acknowledgments}

The help of  M. Kirsanov and A. Toropin  in calculations is greatly appreciated. The work is supported by 
RFBR grant 13-02-00363.

\end{document}